\documentclass[submission,copyright,creativecommons]{eptcs}
\providecommand{\event}{ICLP 2019} 
\usepackage{breakurl}             
\usepackage{underscore}           

\usepackage{color}
\definecolor{label}{rgb}{0.9,0.5,0.2}
\usepackage{float}
\usepackage{graphicx}
\usepackage{courier}
\usepackage{listings}
\def\lppf{{\tt lppf}}

\title{A Rule-Based System for Explainable Donor-Patient Matching in Liver Transplantation}
\author{Felicidad Aguado$^1$
\email{aguado@udc.es}
\and
Pedro Cabalar$^1$
\email{cabalar@udc.es}
\and
Jorge Fandinno$^2$
\email{jorgefandinno@gmail.com}
\and
Brais Mu\~niz$^1$
\email{brais.mcastro@udc.es}
\and
Gilberto P\'erez$^1$
\email{gperez@udc.es}
\and
Francisco Su\'arez$^{3}$
\email{francisco.suarez.lopez@sergas.es}
\and
\institute{$^1$ IRLab, CITIC Research Center\\ University of A Coru\~na, SPAIN}
\institute{$^2$ University of Potsdam, GERMANY}
\institute{$^3$ Digestive Service, Complexo Hospitalario Universitario de A Coru\~na (CHUAC) \\
Instituto de Investigaci\'on Biom\'edica de A Coru\~na (INIBIC), \\
University of A Coru\~na, SPAIN}
}
\def\titlerunning{A Rule-Based System for Explainable Donor-Patient Matching}
\def\authorrunning{F. Aguado et al.}
\begin{document}
\maketitle


\begin{abstract}
In this paper we present \texttt{web-liver}, a rule-based system for decision support in the medical domain, focusing on its application in a liver transplantation unit for implementing policies for donor-patient matching.
The rule-based system is built on top of an interpreter for logic programs with partial functions, called \texttt{lppf}, that extends the paradigm of \emph{Answer Set Programming} (ASP) adding two main features: (1) the inclusion of partial functions  and (2) the computation of causal explanations for the obtained solutions. The final goal of \texttt{web-liver} is assisting the medical experts in the design of new donor-patient matching policies that take into account not only the patient severity but also the transplantation utility. As an example, we illustrate the tool behaviour with a set of rules that implement the utility index called SOFT. 
\end{abstract}


\section{Introduction}


One of the current problems in decision support from Artificial Intelligence systems is the lack of explanations. 
When a system is making decisions in critical contexts and those decisions may have an impact on people's life like in the medical or legal domains, then explanations turn to be crucial, especially if we expect that a domain expert relies on the obtained answers.
One of these situations from the medical domain where explanations have a crucial role is the process of donor-patient matching in an organ transplantation unit. This process starts when a new organ is received and consists in selecting a patient among those included in a waiting list for transplantation. The transplantation unit is expected to follow an objective policy that takes into account medical parameters and is experimentally supported by the existing records, but more importantly, this decision must be easily reproducible and explicable in a comprehensible way for other agents potentially involved, since it may have life-critical consequences at personal, medical and legal levels. Typically, this decision is taken in terms of a set of numerical weights (the impact of weights variation is studied in~\cite{FSS18}).
Although different classification systems based on Artificial Neural Networks (ANNs) are being proposed (see for instance~\cite{RRNN-liver-transplantation-spain} for the case of liver transplantation) their decisions rely on a black box whose behaviour is not explicable in human terms.

In this paper, we present a rule interpreter, \texttt{web-liver}, designed for assisting the medical experts in the donor-patient matching of a liver transplantation unit, using the case scenario from the Digestive Service in the Corunna University Hospital Center (CHUAC), Spain.
The final goal of this tool is providing a rule editor and interpreter that the experts can interactively use to test different policies (sets of rules), checking not only their accuracy but also the explanations provided for the obtained decisions. The most accepted criterion used for donor-patient matching is a measure of the patient's clinical severity (the \emph{Model for End staging Liver Disease}, or MELD index~\cite{MELD}) but this does not take into account other factors such as the ``utility'' of the transplantation (a prediction of potential success), which may obviously depend on the donor's data as well. The medical experts are interested in designing a set of rules that take into account these different factors, combining both experimental data (possibly through rule learning) and explicit representation based on domain knowledge. The purpose of \texttt{web-liver} is providing a friendly environment where the physicians may try different sets of rules and test their impact in terms of the conclusions that the system provides and the explanations associated to those conclusions. To illustrate the behaviour of \texttt{web-liver} we have started by implementing a policy based on the utility index called \emph{Survival Outcome Following liver Transplantation} (SOFT)~\cite{SOFT}.
Implementing the SOFT index was useful in the requirement analysis sessions to obtain guidelines and specifications from the experts when developing the \texttt{web-liver} behaviour and interface. But, at the same time, it was also interesting to check the results of the SOFT index on the available data. Our dataset consists of the 76 transplant cases from years 2009 and 2010, although we are currently working to expand it. For each case we have variables from both the recipient and the donor as well as from the transplantation itself. 
The tool \texttt{web-liver} provides a web interface (written in \texttt{python}) for a logic programming interpreter called \texttt{lppf} (\emph{Logic Programs with Partial Functions}~\cite{Cab11}), an extension of Answer Set Programming (ASP)~\cite{BET11} with two additional features: (1) partial functions; and (2) computation of explanations for the (functional) answer sets. 


The rest of the paper is organised as follows. First, we informally describe the {\tt lppf} interpreter and its input language. Then, we describe the \texttt{web-liver} system using the computation of the SOFT index as a running example. Finally, we briefly comment about related work and conclude the paper.

\section{Logic Programs with Partial Functions}



To describe the syntax of {\tt lppf} we assume some familiarity with ASP. The main addition with respect to ASP rules is the possibility of using partial functions so each function may have assigned some value in an answer set. This value can be Boolean or from another type, and can be directly assigned in rule heads using the explicit assignment operator \texttt{:=} or the assignment \verb+^=+ of default values. As an example, the rules:
\begin{lstlisting}
punish(P) :- drive(P), alcohol(P)>50.
punish(P) :- resist(P).
sentence(P) ^= innocent :- person(P).
sentence(P) := prison :- punish(P).
\end{lstlisting}
specify that either driving with an alcohol ratio greater than 50 or resisting to authority can be punished, that a default sentence is {\tt innocent} and that being punished implies a {\tt prison} sentence. Given the input data
\begin{lstlisting}
person(gabriel).        person(clare).
drive(gabriel).         drive(clare).
alcohol(gabriel):=60.   alcohol(clare):=0.
resist(gabriel).        ~resist(clare).
\end{lstlisting}
(\verb+~+ denotes explicit negation) we obtain the conclusions:
\begin{lstlisting}
Answer:1
punish(gabriel).
sentence(gabriel)=prison.
sentence(clare)=innocent.
\end{lstlisting}
that, as a main feature of \texttt{lppf} can be justified by additional explanations. For instance, if we ask {\tt lppf} to explain the conclusion \texttt{sentence(gabriel)=prison} we get:

\begin{lstlisting}
*sentence(gabriel) = prison
 |-- punish(gabriel)
 |    |-- alcohol(gabriel) = 60
 |    |-- drive(gabriel)

*sentence(gabriel) = prison
 |-- punish(gabriel)
 |    |-- resist(gabriel)
\end{lstlisting}
Since, in a larger program, these default explanations may easily become too verbose, \lppf{} provides several mechanisms to personalize explanations, both in their content and format to be displayed. A first possibility is labeling those rules that we actually want to be traced in explanation trees (forgetting about the rest). Labeling can be done using a label function or a textual description in natural language. As an example of text labels, the following version of the example:
\begin{lstlisting}
drive(gabriel).
alcohol(gabriel):=60.
resist(gabriel).
"%P has driven drunk" ::
	punish(P) :- drive(P), alcohol(P)>50.
"%P has resisted to authority" :: 
	punish(P) :- resist(P).
"%P has been sentenced to %_Value" :: 
	sentence(P) := prison :- punish(P).
\end{lstlisting}
would produce the next explanations for the conclusion \texttt{sentence(gabriel)=prison}:

\begin{lstlisting}
* (*@\textcolor{label}{gabriel has been sentenced to prison} @*)
 |--  (*@\textcolor{label}{gabriel has driven drunk} @*)

* (*@\textcolor{label}{gabriel has been sentenced to prison} @*)
 |--  (*@\textcolor{label}{gabriel has resisted to authority} @*)
\end{lstlisting}
which are much more readable and can be personalized depending on the target user, her profile or her native language. It is also possible to label rules in groups rather than individually. For instance, the expression:
\begin{lstlisting}
#label r :: resist(P).
\end{lstlisting}
has the effect of labeling with the same label {\tt r} every rule whose head function is {\tt resist(P}). Apart from labeling, we can also decide which conclusions must be included in an explanation query. This is done using a special type of rule like in the example below:

\begin{lstlisting}
#explain sentence(P) :- sentence(P)=prison, alcohol(P)>55, ~resist(P).
\end{lstlisting}
This is asking {\tt lppf} to show the explanations for those facts for function {\tt sentence(P)} that satisfy the conditions in the body.


\section{The {\tt liverLP} and {\tt web-liver} systems}


In order to test the use of \texttt{lppf} for donor-patient matching in the transplantation domain, we have created a decision support system that computes the SOFT index over our data. The set of \lppf{} rules for this task receives the name of \texttt{liverLP} and is divided into four modules: \texttt{facts.lppf}, \texttt{constraints.lppf}, \texttt{rules\_value.lppf} and \texttt{liver\_calc.lppf}.
Module \texttt{facts.lppf} stores all the input data collected from the Hospital transplantation records, including donor data, recipient data and also surgery data. As an example, we show part of the input data for case number {\tt 686}:
\begin{lstlisting}[language=Html, frame=single,breaklines=true,showstringspaces=false]

transplant(686).

age(686):=59.            bmi(686):=21.  
p_trans(686):=0.         ~ab_surgery(686).
albumin_fl(686):=400.    ~dialysis(686). 
icu(686).                ~hospital(686).  
meld(686):=7.            ~l_sup(686).
~enceph(686).            ~p_thromb(686).
~ascites(686).           alive(686).
~portal_bleed(686).      donor_age(686):=63.
~vascular_acc(686).      creatinine_fl(686):=120. 
cold_ischemia(686):=340.
\end{lstlisting}

Using the same input data, the medical experts will try different policies for donor-patient matching. In the particular case of the SOFT index policy, these rules are organized as a set of categories. Each category has an associated weight, which will be added to the total score of a transplant if it meets the conditions of that rule. Also, these rules are divided into two groups: P-SOFT rules and SOFT rules. P-SOFT rules are those applicable before the donor allocation and SOFT rules are the rules only applicable when a donor has been already allocated.
In the \texttt{rules\_value.lppf} module all the SOFT categories and their associated risk values are specified. We show some examples below: 
\begin{lstlisting}[language=Html, frame=single,breaklines=true,showstringspaces=false]
psoft_risk(age_60):=4.
psoft_risk(bmi6_35):=2.
psoft_risk(one_previous_transplant):=9.
psoft_risk(two_previous_transpant):=14.
...

soft_risk(portal_bleed_48h_pretransplant):= 6.
soft_risk(donor_age2_gt_60):= 3.
soft_risk(donor_cerebral_vascular_accident):= 2.
soft_risk(donor_creatinine_15):= 2.
soft_risk(donor_age2_10_20):= -2.
...
\end{lstlisting}
The weight of each category is then added depending on the SOFT conditions for each specific risk, as illustrated below for a pair of cases from the following fragment of \texttt{constraints.lppf}:
\begin{lstlisting}[language=Html, frame=single, breaklines=true, showstringspaces=false]
psoft(bmi6_35,P):= psoft_risk(bmi6_35) :- bmi(P)>35.
soft(donor_age2_10_20,P) := soft_risk(donor_age2_10_20) :- 
      donor_age(P) = Age, Age >= 10, Age <= 20.
\end{lstlisting}
This means that the {\tt bmi} weight (2) must be added if {\tt bmi(P)} is above 35, and the donor's age risk (-2) is added if the donor's age is between 10 and 20.
Both \texttt{psoft} and \texttt{soft} functions are assigned a zero default value for the cases that do not meet the conditions of a particular category. 
%
Module \texttt{liver\_calc.lppf} computes the sum of the weights for all risk values in the functions \verb+psoft_cal(P)+ and \verb+soft_cal(P)+. Then, a discrete risk level is associated depending on the interval in which those values are included:
%
%
%
\begin{lstlisting}[language=Html, frame=single, breaklines=true, showstringspaces=false]
soft_level(P) := low            :- soft_cal(P)=X, X<=5.
soft_level(P) := low_moderate   :- soft_cal(P)=X, 6<=X, X<=15.
soft_level(P) := high_moderate  :- soft_cal(P)=X, 16<=X, X<=35.
soft_level(P) := high           :- soft_cal(P)=X, 36<=X, X<=40.
soft_level(P) := futile         :- soft_cal(P)=X, 40<X.
\end{lstlisting}
We also provide textual descriptions to build readable explanations:
\begin{lstlisting}[language=Html, frame=single, breaklines=true, showstringspaces=false]
#label "Risk level of %P is %_Value because SOFT score is %X" :: soft_level(P) :- soft_cal(P)=X.
#explain soft_level(P).
\end{lstlisting}
This is used to generate textual explanations of the following form:
    \begin{lstlisting}
     Answer:1

    * (*@\textcolor{label}{ Risk level of 686 is low because SOFT score is 0 }@*)
     |-- (*@\textcolor{label}{ Activated rules: }@*)
     |    |-- (*@\textcolor{label}{ cold\_ischemia\_0\_6h 	[-3] }@*)
     |    |-- (*@\textcolor{label}{ donor\_age2\_gt\_60 	[3] }@*)
     
     (. . .)
    
    * (*@\textcolor{label}{ Risk level of 763 is high\_moderate because SOFT score is 22 }@*)
     |-- (*@\textcolor{label}{ Activated rules: }@*)
     |    |-- (*@\textcolor{label}{ donor\_cerebral\_vascular\_accident 	[2] }@*)
     |    |-- (*@\textcolor{label}{ psoft 	[20] }@*)
     |    |    |-- (*@\textcolor{label}{ intensive\_care\_unit\_pretransplant 	[6] }@*)
     |    |    |-- (*@\textcolor{label}{ life\_support\_pretransplant 	[9] }@*)
     |    |    |-- (*@\textcolor{label}{ portal\_vein\_thrombosis 	[5] }@*)
    
    76 ocurrences explained.
    
    1 solution
    \end{lstlisting}
Although the \texttt{liverlp} system is convenient for fast prototyping, it is far from being directly usable by medical specialists who are not familiar with logic programming languages. For this reason, we have built the \texttt{web-liver} system, a web application written in python that allows creating and manipulating rules through a web interface, and testing their behaviour on the records data, displaying the obtained conclusions and their explanations in a more visual way. Its functionality is divided in three modules: \texttt{classifiers}, \texttt{results} and \texttt{transplants}. 
The first one, \texttt{classifiers}, allows the user to create, modify and delete different rule groups which are called classifiers (being the SOFT classifier one among them). The user can add, delete and configure the set of rules of each classifier. For each rule, she can change its \texttt{lppf} label, its value and its conditions set (the rule body). An example from the rule edition window is shown below:
\begin{center}
\includegraphics[scale=0.36]{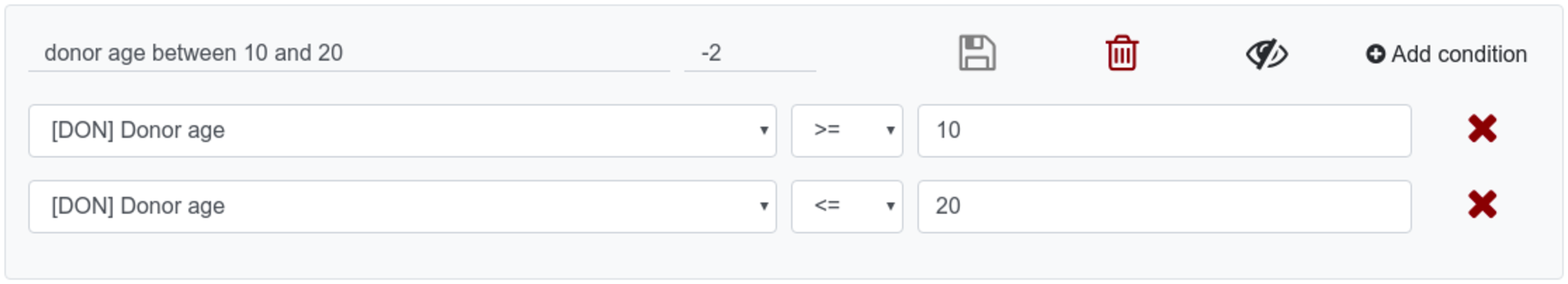} 
\end{center}
This will automatically generate the \lppf{} code:
\begin{center}
    \texttt{"donor age between 10 and 20" :: rule(P):=-2 :- donor\_age(P)>=10, donor\_age(P)<=20.}
\end{center}
The tool allows creating new classifiers or copying them from previously existing ones, so that, for instance, it is easy to create a small modification of the SOFT index, adding or changing some of its rules and/or weights.
Once the user selects a classifier, \texttt{web-liver} will build, solve and show the explanations for the conclusions obtained from the existing data base. Apart from textual explanations as the ones shown before, \texttt{web-liver} also allows generation of graphs like the one in the following picture:
\begin{center}
        \includegraphics[width=1\textwidth]{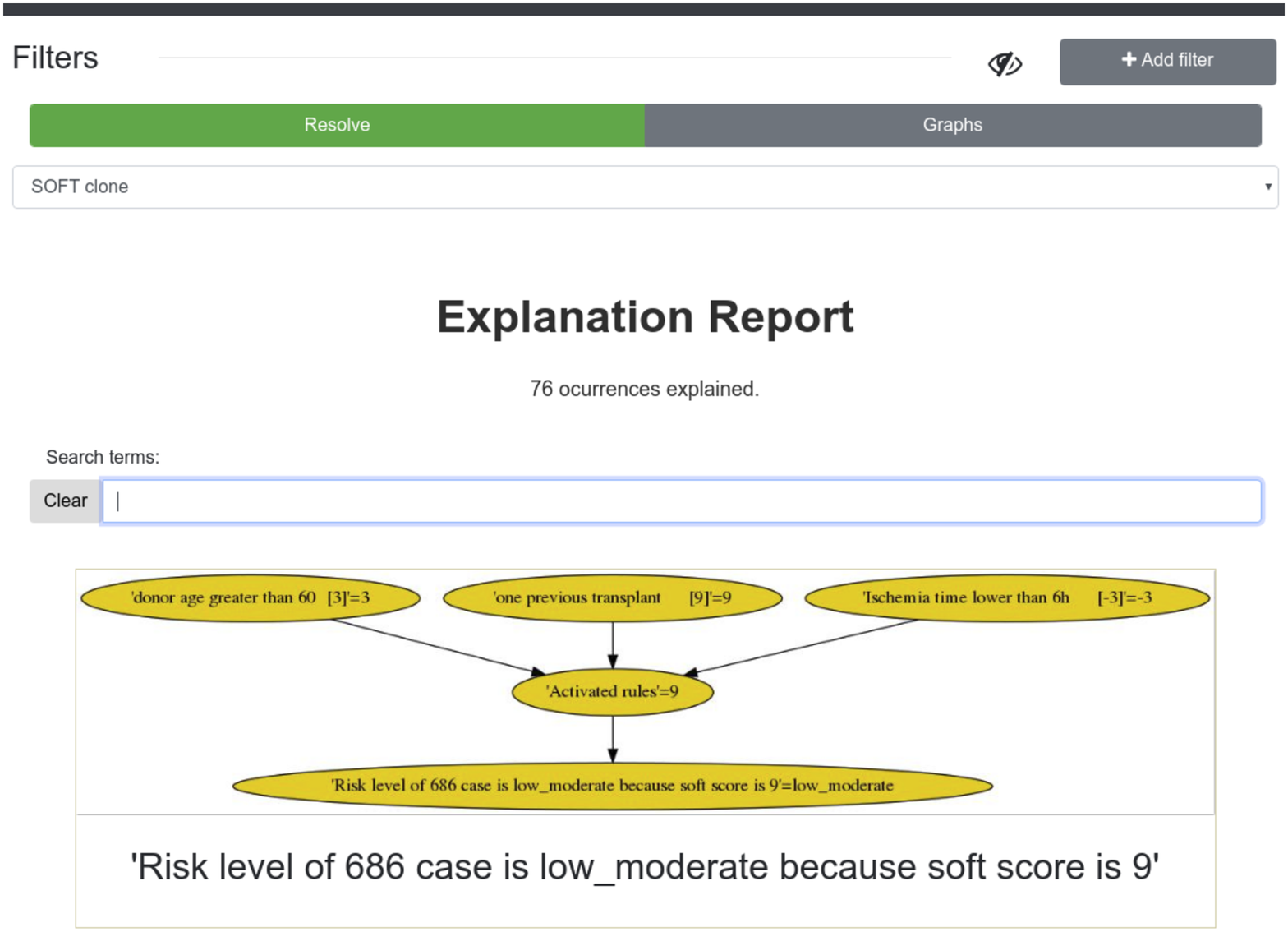} 
\end{center}
These graphs are grouped in an HTML report that can be additionally filtered in the web interface. Finally, the \texttt{transplants} module allows the user to explore the transplantation cases data one by one and check the results of applying a classifier to a concrete case.

\section{Related work and conclusions}

We have built a support tool for editing rules, interpreting them and explaining their results in an environment for donor-patient matching in liver transplantation. The selection of a set rules and their adequacy from a medical perspective is out of the scope of the current paper. The final goal is obtaining an explainable classifier that can be designed as a combination of expert knowledge or learning methods (for instance to adjust the classifier weights) using ANNs as in~\cite{RRNN-liver-transplantation-spain} or random forest tress as in~\cite{Forest-trees-lau}.
From a logic programming perspective, the explanation features from \lppf{} are based on~\cite{Cgraph} but there exist other approaches for justification of ASP programs (see~\cite{FS19} for a recent survey). 
Similarly, the functional extension of ASP is based on~\cite{Cab11} since, although other functional extensions exist, it is the only one allowing free nesting of functional terms, a feature that can be freely used in any \lppf{} rule. 
For future work, we plan to keep extending the \lppf{} language with new aggregate functions and with causal literals \cite{CausalLit} that would allow rule conditions that test if an atom has been a cause of another atom. For the {\tt liverLP} system, our plans include extending the dataset and applying (symbolic) learning algorithms or integration with public online medical ontologies.


\appendix


\bibliographystyle{eptcs}
\bibliography{refs}

\appendix

\end{document}